\documentclass[review]{elsarticle}

\usepackage{lineno,hyperref}
\usepackage{amsmath} 
\usepackage[version=4]{mhchem}		
\usepackage{color}								
\usepackage[full]{textcomp}

\graphicspath{{./Figures/}}

\journal{Journal of Quantative Spectroscopy \& Radiative Transfer}

\begin{document}

\begin{frontmatter}

\title{Circular spectropolarimetric sensing of chiral photosystems in decaying leaves}


\author[mymainaddress]{C.H. Lucas Patty}\corref{mycorrespondingauthor}
\cortext[mycorrespondingauthor]{Corresponding author}
\ead{lucas.patty@vu.nl}

\author[a]{Luuk J.J. Visser}
\author[b]{Freek Ariese}
\author[c]{Wybren Jan Buma}
\author[d]{William B. Sparks}
\author[mymainaddress]{Rob J.M. van Spanning}
\author[mymainaddress]{Wilfred F.M. R{\"o}ling \textsuperscript{\textdied}}
\author[a]{Frans Snik}

\address[mymainaddress]{Molecular Cell Biology, Amsterdam Institute for Molecules, Medicines and Systems, VU Amsterdam, De Boelelaan 1108, 1081 HZ Amsterdam, The Netherlands}
\address[a]{Leiden Observatory, Leiden University, P.O. Box 9513, 2300 RA Leiden, The Netherlands}
\address[b] {LaserLaB, VU Amsterdam, De Boelelaan 1083, 1081 HV Amsterdam, The Netherlands}
\address[c] {HIMS, Photonics group, University of Amsterdam, Science Park 904, 1098 XH Amsterdam, The Netherlands}
\address[d] {Space Telescope Science Institute, 3700 San Martin Drive, Baltimore, MD 21218, USA}

\begin{abstract}
Circular polarization spectroscopy has proven to be an indispensable tool in photosynthesis research and (bio)-molecular research in general. Oxygenic photosystems typically display an asymmetric Cotton effect around the chlorophyll absorbance maximum with a signal $\leq 1 \%$. In vegetation, these signals are the direct result of the chirality of the supramolecular aggregates. The circular polarization is thus directly influenced by the composition and architecture of the photosynthetic macrodomains, and is thereby linked to photosynthetic functioning. Although ordinarily measured only on a molecular level, we have developed a new spectropolarimetric instrument, TreePol, that allows for both laboratory and in-the-field measurements. Through spectral multiplexing, TreePol is capable of fast measurements with a sensitivity of $\sim 1*10^{-4}$ and is therefore suitable of non-destructively probing the molecular architecture of whole plant leaves. We have measured the chiroptical evolution of \textit{Hedera helix} leaves for a period of 22 days. Spectrally resolved circular polarization measurements (450-900 nm) on whole leaves in transmission exhibit a strong decrease in the polarization signal over time after plucking, which we accredit to the deterioration of chiral macro-aggregates. Chlorophyll \textit{a} levels measured over the same period by means of UV-Vis absorption and fluorescence spectroscopy showed a much smaller decrease. With these results we are able to distinguish healthy from deteriorating leaves. Hereby we indicate the potency of circular polarization spectroscopy on whole and intact leaves as a nondestructive tool for structural and plant stress assessment. Additionally, we underline the establishment of circular polarization signals as remotely accessible means of detecting the presence of extraterrestrial life.

\end{abstract}

\begin{keyword}
\texttt{Circular polarization  \sep Photosynthesis \sep Biosignatures \sep Homochirality \sep Life Detection}

\end{keyword}

\end{frontmatter}


\section{Introduction}

Terrestrial biochemistry is based upon chiral molecules, which largely determine the functioning and structure of biological systems. Amino acids primarily occur in the L-configuration while sugars occur predominantly in the D-configuration. Biological macromolecules are often also chiral. The $\alpha$-helix, a common secondary structure of proteins, for example is almost exclusively right-hand-coiled. Homochirality is a prerequisite for self-replication and thus terrestrial life \cite{Popa2004, Bonner1995, Jafarpour2015}. It is highly likely that homochirality is a universal feature of life. Analysis of various chondrite samples has shown that they contain slight enantiomeric excesses \cite{Cronin1997, Pizzarello2000} and recently chiral molecules were discovered in interstellar molecular clouds \cite{McGuire2016}. Furthermore, it has been proposed that circular polarization as a result of UV scattering can generate slight enantiomeric excesses in protosolar systems \cite{Bailey1998}, which can seed the chiral evolution of life. 

The phenomenon of chirality causes molecules to interact with polarized light in a number of different ways. Circular dichroism, for example, is a measure of the differential absorbance of left and right handed circularly polarized light $(\Delta A)$. 

For noncoupled chromophores (for instance pigments in a solution that does not allow interaction between them), the shape of the circular dichroism spectra and the absorption bands are similar. By contrast, multiple chirally oriented chromophores, even when achiral themselves, will exhibit an exciton circular dichroism spectrum. The exciton spectrum is characterized by two bands of opposite sign around the wavelength of maximum absorbance, which exhibits no circular polarization preference \cite{Hembury2008,Gottarelli2008,Berova2000}. While circular dichroism spectroscopy offers a quick insight into the architecture of these molecular systems, it is often difficult to unravel the exact structures based on the spectrum alone. Therefore, careful comparison with models and results from controlled experiments are required to take full advantage of the measurements \cite{Hembury2008,Berova2000}. 

\begin{figure*}[ht]
  \centering
  \includegraphics[width=\textwidth]{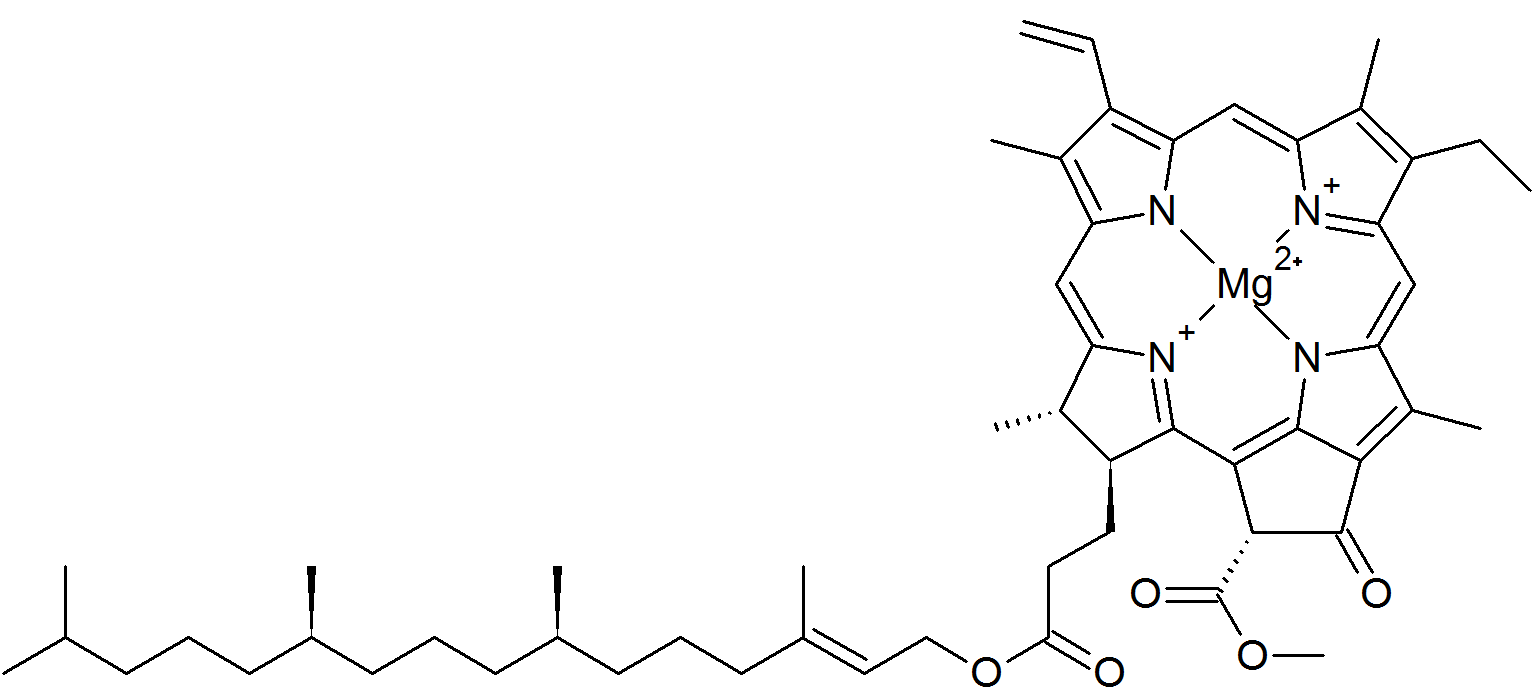}
  \caption{The structure of chlorophyll \textit{a}}
  \label{fig:Chlorophyll}
\end{figure*}

Chirality is also observed in chlorophylls and bacteriochlorophylls. Oxygenic photosynthesis, such as occurs in plants, algae and cyanobacteria, is a process during which light energy is converted into chemical energy, which in turn is used for the assimilation of \ce{CO2} into organic matter. During that process, \ce{O2} is also produced. Photosynthesis is the major driving force of life on Earth and evolved soon after the emergence of life on Earth \cite{Hohmann-Marriott2011,Blankenship2010,Xiong2002}. Likely, given that life has access to light, the development of photosynthesis is a universal phenomenon \cite{Wolstencroft2002,Rothschild2008}. Chlorophylls are the most important cofactors in photosynthesis and are cyclic tetrapyrroles with a central Mg, a characteristic isocyclic five-membered ring and a long-chain esterifying alcohol at \ce{C-17} \cite{Scheer2006}. Chlorophylls thus are inherently chiral (albeit with a very weak intrinsic circular polarization signal). One of the most common chlorophylls in vegetation, chlorophyll \textit{a}, is shown in Figure \ref{fig:Chlorophyll}. In vegetation, the chlorophylls are furthermore organized in chiral protein-pigment aggregates. Similar to other chiral aggregates, it is the supramolecular structure, the large-scale handedness, that dominates the polarization spectrum rather than the chirality of the constituent molecules themselves (although these signals are to some degree superimposed, the total spectrum is not the sum of the spectra of the constituents alone) \cite{Keller1986, Barzda1994}. Such chiral aggregates can cause very intense polarization signals that may be two orders of magnitude larger than the molecular signal with non-conservative anomalously shaped bands that range beyond the wavelengths of molecular absorbance (Qy at 680 nm for chlorophyll \textit{a} in water) \cite{Garab2009}. In the case of large-scale aggregates, such as found in photosynthesis, the resultant signal is not only influenced by differential absorption but also by differential scattering \cite{Bustamante1983}. As the circular polarization signal is dependent on the interactions on a molecular scale, circular polarization measurements are a unique way of probing the molecular \textit{organization} remotely.

Recently, it was shown that the  circular polarization by phototrophic organisms can be successfully measured \textit{in vivo} \cite{Sparks2009a, Sparks2009,Toth2016}. Sparks et al. \cite{Sparks2009a, Sparks2009} successfully measured the circular polarization of fresh leaves and phototrophic bacteria, both in reflectance and transmittance, using diffuse unpolarized incident light. T{\'o}th et al. \cite{Toth2016} measured the \textit{in vivo} circular dichroism of water-infiltrated leaves of various plant species and \textit{Arabidopsis thaliana} mutants using a commercially available circular dichroism spectrometer. Although \textit{in vivo} measurements are still quite rare and mainly used as benchmark observations for astrobiology, circular polarization has the potential to provide a clear and unambiguous biosignature. Remotely detectable spectral characteristics have mainly focused on detecting particular atmospheric constituents. Such constituents include \ce{H2O}, which indicates possible planetary habitability, but also gases that could result directly or indirectly from biological activity such as \ce{O2} and \ce{CH4} \cite{Kaltenegger2007, DesMarais2002}. Detection of neither of these gases is, however, free of false-positive scenarios \cite{Domagal-Goldman2014, Schwieterman2016, Harman2015, Wordsworth2014}. Other suggested remotely detectable biosignatures include the vegetation red edge \cite{Seager2005, Kiang2007} or pigment signatures by non-phototrophs \cite{Schwieterman2015}, but there is a potential risk of possible false-positives by mineral reflectance. 

It has been suggested that linear polarization signatures may also be good potential biosignatures \cite{Berdyugina2016}. Berdyugina et al. measured the linear polarization signal of various leaves at near Brewster's angle and found that the combination of reflectance and linear polarimetric signals allows to distinguish between abiotic and biotic materials. Both biotic and abiotic matter, however, can create linear polarization \cite{Shkuratov2006, West1997}.  

Circular polarization is more exclusive and signatures of larger amplitudes require larger order dissymmetry, something particularly found in nature. Laboratory measurements on various minerals consistently showed a much weaker signal and different in shape \cite{Sparks2009, Pospergelis1969} and the circular polarization imaging of the surface of Mars did not find any significant signals, attesting to a general lack of false positives \cite{Sparks2005}. Furthermore, while minerals can be chiral, an enantiomeric excess on a planetary scale would be required.  

Additionally, circular polarization could be utilized in the remote sensing of vegetation on Earth. From the viewpoint of remote sensing, it has been suggested that the scalar regular reflectance spectra contain more and better features than polarization spectra \cite{Peltoniemi2015}. The structural and organizational information on the photosystems provided by circular  polarization, however, is unique and could prove to be a valuable tool in assessing vegetation physiology on Earth.

In the present study we show for the first time, through spectral multiplexing, how the circular polarization signal of a leaf in transmission decays over a period of 22 days after detachment. These results allow us to distinguish between healthy and deteriorating matter, which would potentially allow to monitor vegetation stress in greater detail. With these results we also want to emphasize the potency of circular polarization as an unambiguous biosignature.  

While in this paper we only show measurements in transmission, we designed and built the TreePol instrument in order to also carry out circular spectropolarimetric reflectance measurements both in the laboratory and in the field. We have already started employing our instrument for extensive spectropolarimetric reflectance studies. The polarization signal in transmission is, however, an order of magnitude larger than in reflectance, which allows us to assess the signal decrease in greater detail. The reflectance results will be presented in a subsequent paper, geared towards the use of circular polarization in remote sensing.


\section{Materials and Methods}
\subsection{Sample collection and storage}
Fresh leaves of \textit{Hedera helix} (common ivy; juvenile phenotype), petiole removed, were collected in June from a private backyard garden near the city centre of Amsterdam, rinsed with distilled water and padded dry. We have stored the leaves at room temperature under two conditions: one set of leaves (number of leaves; n=6) was stored in the dark and one set of leaves (number of leaves; n=6) was stored at a 12 h/day light regime (at 80 $\mu$moles photons /m$^{2}$/s photosynthetically active radiation (400-700 nm)) to simulate daylight. The spectral properties of the leaves were measured at 0, 2, 4, 7, 10, 14 and 22 days and we weighed the leaves before every measurement. 
The leaves used for the circular dichroism measurements were collected from the same plant in December. 

\begin{figure*}[ht]
  \centering
  \includegraphics[width=\textwidth]{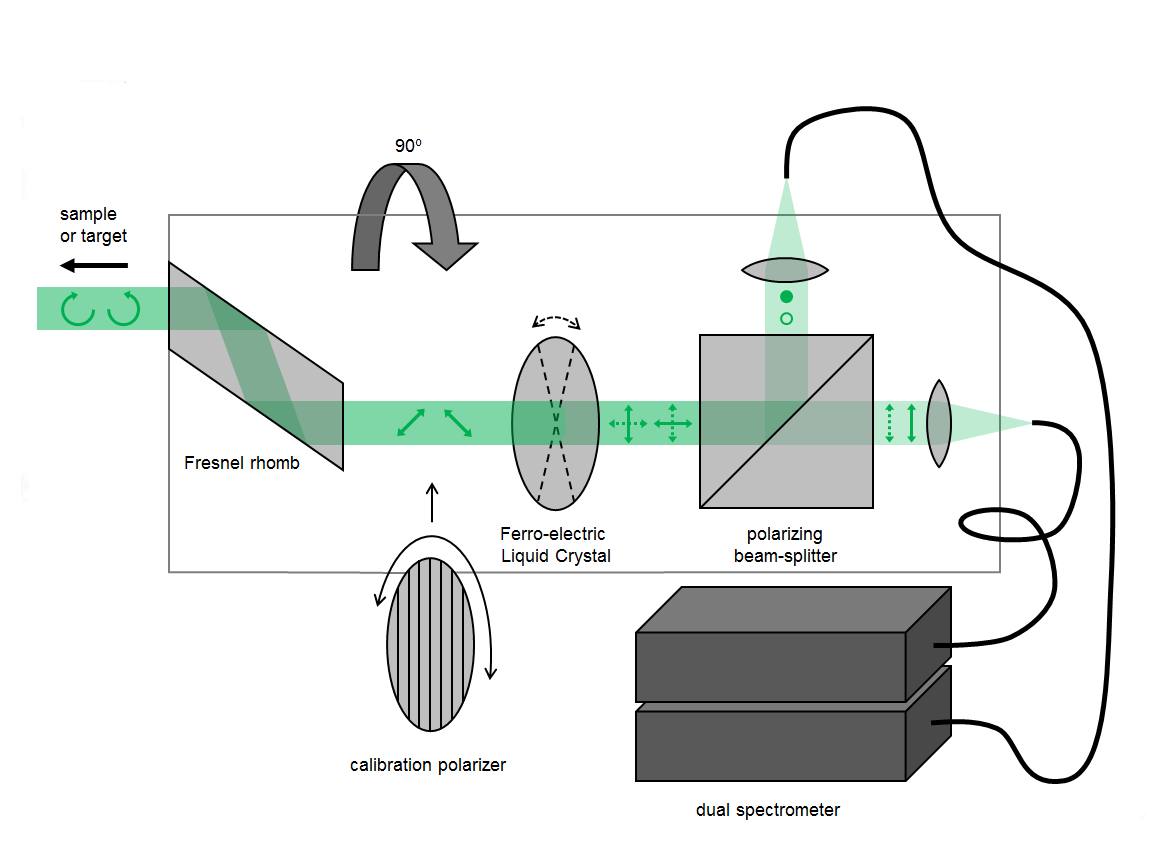}
  \caption{Schematic presentation of TreePol.}
  \label{fig:TreePol}
\end{figure*}

\begin{figure*}[h]
  \centering
  \includegraphics[width=\linewidth]{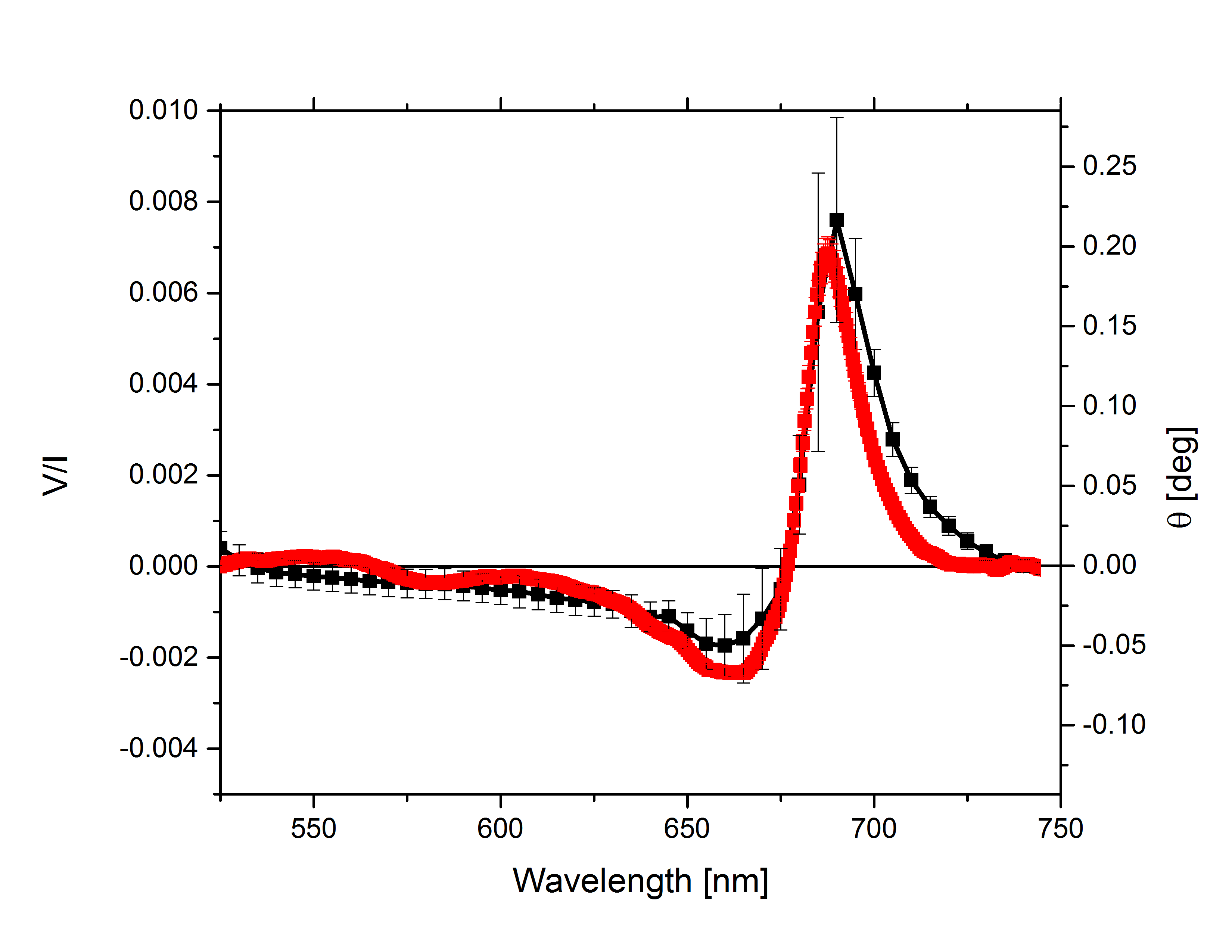}
  \caption{Comparison of TreePol (Red, n=12) and Chirascan (Black, n=4) spectra for two representative sets of fresh \textit{H.helix} leaves. Error bars denote the SE. The Chirascan readout was converted to $V/I$ units as described in the text, $\theta$ is displayed for clarity}
  \label{fig:TreePolCD}
\end{figure*}

\subsection{Circular polarization and circular dichroism} \label{sec:CPCD}
Circular dichroism is traditionally often expressed in degrees of ellipticity $(\theta)$, where the ratio of the minor to the major axis of the resultant polarization ellipse defines the tangent of the ellipticity \cite{Fasman2013}. In fields other than biology and biochemistry, circular polarization and polarization in general is often described in terms of the four parameters of the Stokes vector $\textbf{S}$. With the electric field vectors $E_{x}$ in the x direction ($0^{\circ}$) and $E_{y}$ in the y direction ($90^{\circ}$), the Stokes vector is given by: 

\begin{equation*}
\textbf{S}=
\begin{pmatrix}
    I\\
    Q\\
    U\\
    V\\
\end{pmatrix}=
\begin{pmatrix}
    \left\langle E^{}_{x}E^{*}_{x} + E^{}_{y}E^{*}_{y}\right\rangle\\
    \left\langle E^{}_{x}E^{*}_{x} - E^{}_{y}E^{*}_{y}\right\rangle\\
    \left\langle E^{}_{x}E^{*}_{y} - E^{}_{y}E^{*}_{x}\right\rangle\\
    i\left\langle E^{}_{x}E^{*}_{y} - E^{}_{y}E^{*}_{x}\right\rangle\\
\end{pmatrix}=
\begin{pmatrix}
    I_{0^{\circ}}+I_{90^{\circ}}\\
    I_{0^{\circ}}-I_{90^{\circ}}\\
    I_{45^{\circ}}-I_{-45^{\circ}}\\
    I_{RHC}-I_{LHC}\\
\end{pmatrix}
\end{equation*}

The Stokes parameters $I$, $Q$, $U$ and $V$ thus refer to electric field intensities, which thereby relate to measurable quantities. $I$ is the absolute intensity, $Q$ and $U$ denote linear polarization, with $Q$ the difference between horizontal and vertical polarization. Similarly, $U$ denotes linear polarization but with + $45^{\circ}/-45^{\circ}$. Finally, $V$ gives the difference between left-handed circularly polarized light and right-handed circularly polarized light. If we thus know the absolute intensity $I$, the polarization state can be completely described by $Q/I$, $U/I$ and $V/I$. In the ideal case of a completely circularly polarizing sample $V/I$ will be -1 for left-handed circularly polarized light and 1 for right-handed circularly polarized light. $I_{0^{\circ}}, I_{90^{\circ}}, I_{45^{\circ}}$ and $I_{-45^{\circ}}$ are the intensities oriented in the planes perpendicular to the propagation axis and $I_{LHC}$ and $I_{RHC}$ are, respectively, the intensities of left- and right-handed circularly polarized light.  Additionally, it is required that the incoming light, i.e., the light interacting with the sample, is unpolarized itself. While $V/I$, $\theta$ and $\Delta A$ are measured in a different way, the three are comparable, however, by $\theta = \Delta A * ln(10)/4 * (180/\pi)$, $V/I = \Delta A * ln(10)/2$ and $\theta = (V/I)/2 * (180/\pi)$. This relation results from: $\theta (rad) = (\sqrt{I_{r}}-\sqrt{I_{l}})/(\sqrt{I_{r}}+\sqrt{I_{l}})$ and $V/I =  (I_{r}-I_{l})/(I_{r}+I_{l})$. $\Delta A$ can then be obtained by substituting Lambert-Beer for $I$.

\subsection{TreePol}
TreePol is a spectropolarimetric instrument developed by the Astronomical Instrumentation Group of the Leiden Observatory (Leiden University). In contrast to existing commercial circular dichrographs which modulate light before interaction with the sample (and thus measure the response of a sample to circularly polarized light), Treepol measures the fractional circular polarization of light after interaction of the sample with unpolarized light. Furthermore, in contrast to existing commercial circular dichrographs \cite{Greenfield2006} (which are all based on a photoelastic modulator and a main scanning monochromator, and are thus very slow) or other sensitive circular spectropolarimetric equipment \cite{Sparks2009, Sparks2009a}, TreePol applies spectral multiplexing with the implementation of a dual fiber-fed spectrometer. The polarimetric sensitivity is obtained by using ferro-liquid-crystal (FLC) modulation synchronized with fast read-out of the one-dimensional detector in each spectrograph, in combination with a dual-beam approach in which a polarizing beam splitter feeds the two spectrographs with orthogonally polarized light. This combination of temporal polarization modulation (i.e.~the combination of a fast FLC, which can flip its fast axis by speeds up to multiple kHz, and a high-speed spectrograph) with spatial modulation (i.e.~simultaneous recording of orthogonal polarization states using two synchronized spectrographs) in a so-called 'beam-exchange' ensures that systematic differential effects are canceled out (to first order) and do not induce spurious polarization signals down to the $\sim 10^{-5}$ level \cite{Snik2013}. The measurement efficiency and insensitivity to temporal effects (like swaying trees) would render TreePol a highly suitable instrument for remote-sensing observations in the field, which is one of the ultimate goals of this research line. 

TreePol was specifically developed to measure the fractional circular polarization as a function of wavelength (400-900 nm) with an accuracy of $<1*10^{-3}$, a sensitivity of $1*10^{-5}$, a maximal efficiency of 95\% and a maximal frame rate of 952 Hz. TreePol is capable of measuring the circular polarization of a sample in both transmittance and reflectance, depending on the angle of the light source (KL1500, Schott AG, Germany). After interaction with the sample, the Fresnel rhomb (FR600QM, Thorlabs, USA) converts (with an efficiency of $>99\%$) the measurable circular polarization into linear polarization states at $\pm$45$^\circ$ that can be modulated by the FLC, which is a half-wave retarder at 590 nm (LV1300-OEM liquid crystal polarization rotator, Micron Technology, USA). By controlling the voltage upon the FLC (using the associated electronics box and custom-built synchronization electronics) to $\pm$$\sim$5 V (with a duty cycle such that the average voltage is zero), the half wave plate's fast axis flips between $\pm$22.5$^\circ$. In this way, the relevant polarization states are converted to the (horizontal/vertical) polarization states that are split by the polarizing beam splitter (BV-100-VIS, Meadowlark Optics, USA). The FLC modulation continuously swaps the polarization content of the two beams, which is converted to intensity signals in the spectrographs (AvaSpec 1650F-2-USB2, Avantes, The Netherlands) that scale with $[I\pm V](\lambda)$ for one beam, and $[I\mp V](\lambda)$ for the other. By combining the four redundant sets of intensity spectra, the circular polarization spectrum $V/I$ is obtained after a double-difference scheme \cite{Snik2013}. A diagram of the instrument is shown in Figure \ref{fig:TreePol}.

Calibration of TreePol is fast and straightforward. With the Fresnel rhomb placed parallel to the s direction of the polarized beam splitter, the fast axis is positioned at $+U$. A linear polarizer (GT-10, Thorlabs, USA) is then positioned between the Fresnel rhomb and the FLC at $\pm U$. The FLC can be positioned accurately at $\pm Q$, using real-time spectrometer data streaming (while correcting for measurement data without the linear polarizer and FLC in place). With the FLC aligned a measurement is taken without the linear polarizer. Thereafter a measurement is taken with the linear polarizer in place positioned at $\pm Q$. Using these two measurements the FLC efficiency is calculated, which consequently is corrected for the Fresnel rhomb retardance and the extinction of the polarizing beam splitter to retrieve the total efficiency for $V$.

FLCs have proven to be a valid alternative for photoelastic modulators \cite{Gandorfer1999}, and have been employed in other sensitve spectropolarimetric instruments \cite{Bailey2015}. FLCs, however, are known to generate wavelength dependent polarized fringes which are influenced by temperature \cite{JuanOvelar2012}. From 525 to 750 nm these fringes have a maximum value of $1*10^{-2}$ but can be calibrated out. Variations in ambient conditions during measurements, however, can shift the wavelength dependency of the fringes. In this study, leftover fringes may have a maximum amplitude of $5*10^{-4}$ from 525 to 750 nm, but reduce in amplitude with increasing wavelength. From 650 to 750 nm the maximum amplitude is $<2*10^{-4}$. Errors due to fringes are separately displayed in the results.

Due to imperfections, mostly of the FLC properties (retardance offset from half-wave away from the central wavelength, polarized spectral fringing, misalignment), and cross-talk by the Fresnel rhomb, the instantaneous $V/I$ measurement is susceptible to linear polarization of the incident light, which is often much stronger than the circular polarization signals. By rotating the entire instrument (which is positioned in a rotating inner cage assembly and can therefore be easily rotated by hand) by 90$^\circ$, the sign changes on any linear polarization signal, whereas the circular polarization is invariant for rotation. A second set of measurements can therefore be used to obtain through a 'triple-difference' yield a measure of $V/I$ that is virtually free of linear polarization cross-talk, as long as the rotation is exactly 90$^\circ$, and the sample/target and/or observing conditions are unchanged. 

All leaves were measured with illumination of the adaxial side. No corrections for leaf geometry were carried out and an area with radius $r \approx 1$ cm was measured. All measurements were performed with an integration time of 18 ms and a 180 seconds total measurement duration for each full spectral scan.       

\subsection{Circular dichroism spectropolarimetry} 
In vivo circular dichroism (i.e. in transmission) was additionally measured using a commercially available CD dichrograph (Chirascan Plus, Applied Photophysics, UK). Small strips of leaf (n=6) were placed in a quartz cuvette (10 mm x 10 mm) with a small metal placeholder. The adaxial side was illuminated. Spectra were recorded from 400 to 800 nm, using a 5 nm band pass and a 5 nm step size. All measurements were repeated 3 times with a 5 s time-per-point and were carried out at room temperature. 

\subsection{Extraction and absorbance/fluorescence spectroscopy}
119 mm$^{2}$ of leaf surface was frozen in liquid nitrogen and disrupted mechanically into a fine powder using quartz beads (using three different leaves, n=3). We extracted the chlorophylls by adding 1 ml cold neutralized methanol and incubated the mixture for 90 minutes. Hereupon we centrifuged the solution at 20000 g for 5 minutes and discarded the pellet. 
Absorbance measurements, where $A=-\log(I/I_{0})$, were performed using a Cary 50 (Varian, USA) UV-Vis spectrophotometer in 10 mm x 10 mm quartz cuvettes from 300 to 800 nm. For absorbance measurements we diluted the chlorophyll extract 5 times with methanol. Fluorescence measurements were performed using a Cary Eclipse (Varian, USA) fluorescence spectrometer. Excitation and emission monochromator slit widths were set at 10 nm and 5 nm, respectively, and 10 mm x 10 mm quartz cuvettes were used. The samples were excited at 434 nm and emission was measured from 600 to 800 nm. For fluorescence measurements we diluted the chlorophyll extract 500 times with methanol in order to minimize self-absorption. Measurements were carried out at room temperature. In order to quantify the amount of chlorophyll \textit{a} we compared the sample data with a calibration series of essentially pure chlorophyll \textit{a} (Sigma Aldrich, USA) in methanol under identical conditions. We have exclusively quantified the concentrations of chlorophyll \textit{a}, which is the main pigment in vegetation. Spectroscopic analysis showed that chlorophyll \textit{b} was present in much lower quantities and was very stable over time.

\section{Results}

\begin{figure*}[]
  \centering
  \includegraphics[height=\textheight]{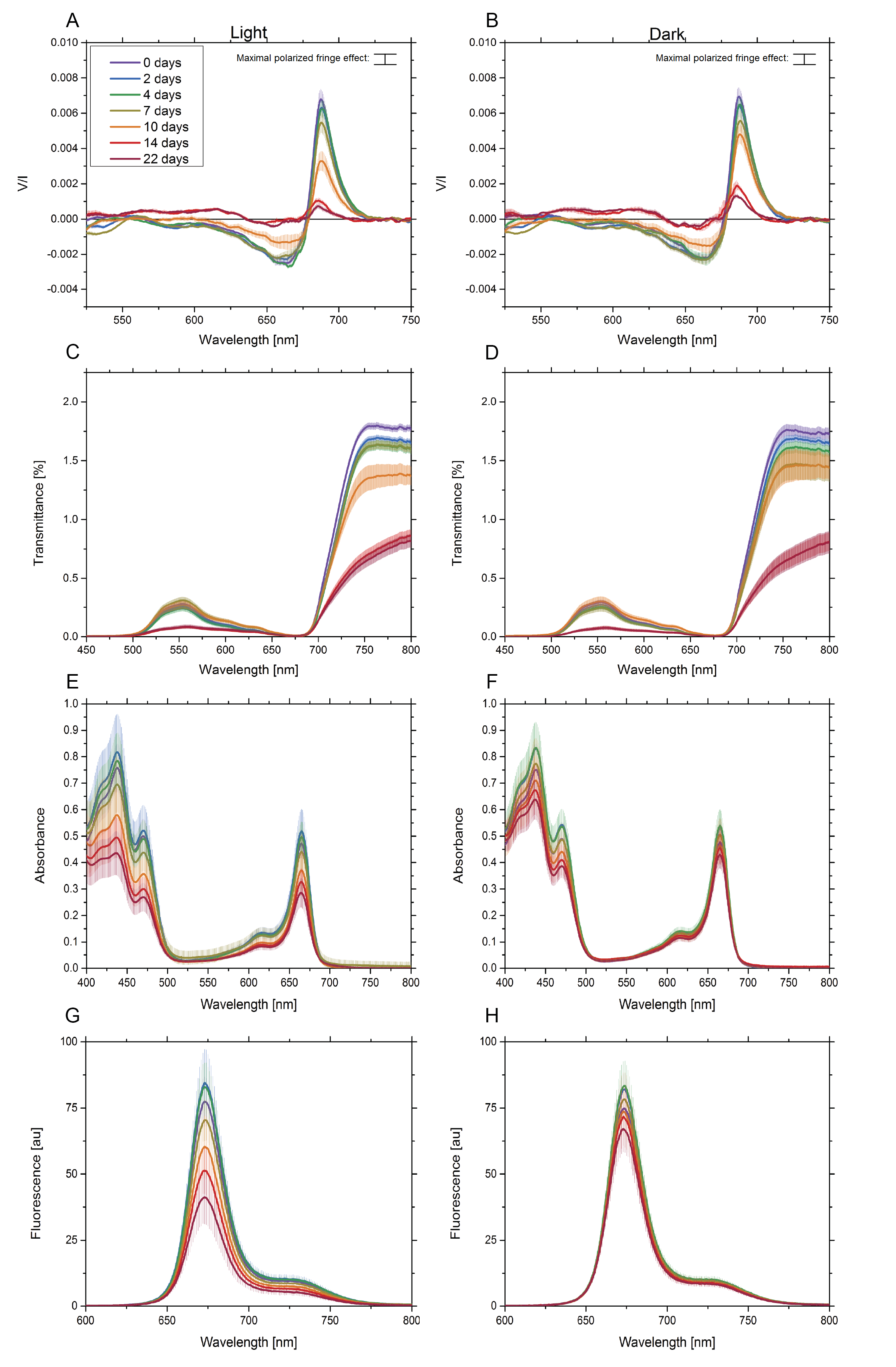}
  \caption{\textbf{(A)} and \textbf{B} show TreePol V/I over time of the daylight and dark stored leaves.\textbf{(C)} and \textbf{(D)} show TreePol transmittance over time of the daylight and dark stored leaves. Error bars for \textbf{A}, \textbf{B}, \textbf{C} and \textbf{D} denote the SE, n=6. \textbf{(E)} and \textbf{(F)} show the absorbance over time of the chlorophyll extracts of the daylight and dark stored leaves. \textbf{(G)} and \textbf{(H)} show the fluorescence over time of the chlorophyll extracts of the daylight and dark stored leaves. Error bars for \textbf{E}, \textbf{F}, \textbf{G} and \textbf{H} denote the SD, n=3.}
  \label{fig:Spec}
\end{figure*}

\begin{figure*}[]
  \centering
  \includegraphics[width=\linewidth/2]{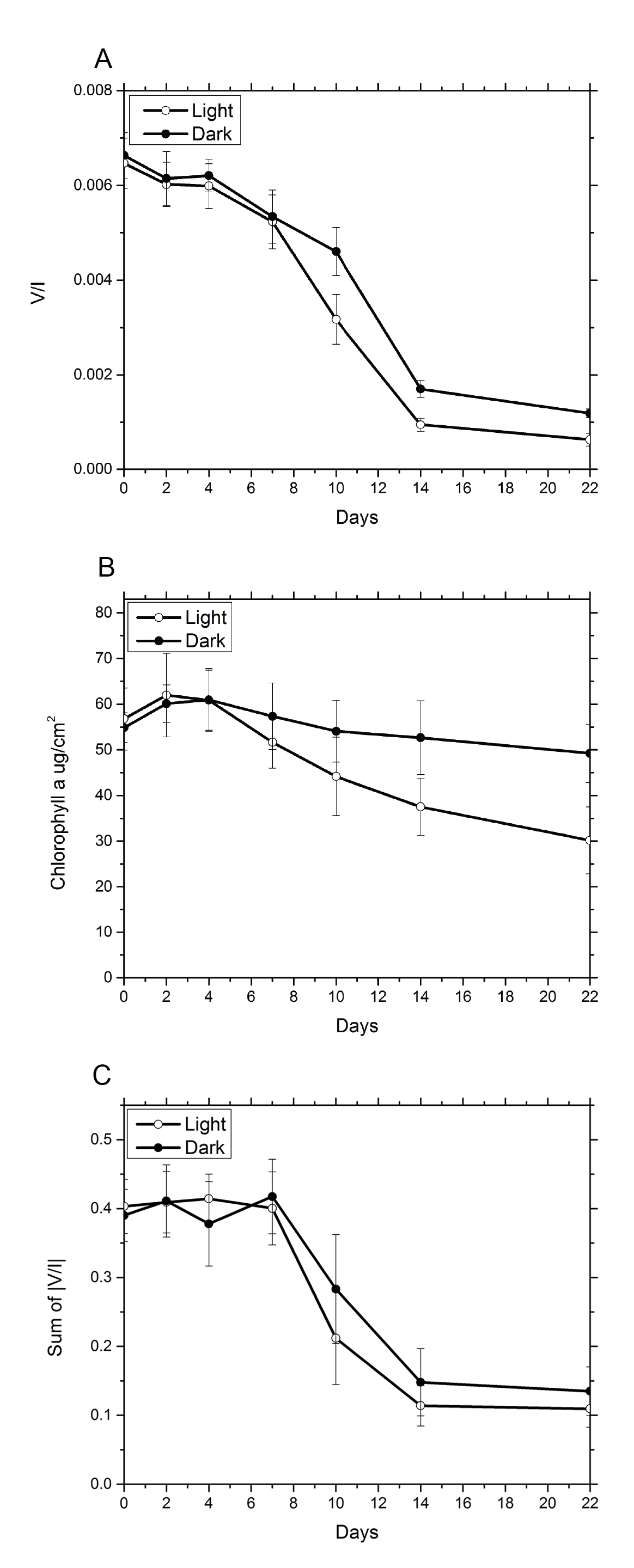}
  \caption{\textbf{(A)} Average of the $V/I$ peak at 685-692 nm over time for the daylight and dark stored leaves. \textbf{(B)} Chlorophyll \textit{a} concentrations per cm$^{2}$ over time for the daylight and dark stored leaves. \textbf{(C)} Sum of  absolute $V/I$ over 525-750 nm. Error bars denote the SE (n=6) for \textbf{(A)} and \textbf{C} and SD (n=3) for \textbf{(B)}.}
  \label{fig:Day}
\end{figure*}

\subsection{Dark/Daylight differences and weight loss}
We determined the weight loss of the leaves before every measurement and observed a gradual decrease over time, mainly due to water loss (confirmed by IR spectroscopy, results not shown). Importantly, the mean difference in relative weight loss between the leaves stored permanently in the dark and those stored at simulated daylight was always smaller than 0.93\%. 

\subsection{Circular Polarization}
As shown in figure \ref{fig:TreePolCD} \textbf{A}, the results from both TreePol and the commercial circular dichrograph show a signal of similar shape and amplitude. It is important to realize that the two instruments employ two quite different approaches to measure the circular polarization spectrum. Circular dichroism spectroscopy is based on the modulation of light (in either left- or right-handed circularly polarized light) before it interacts with a sample and the difference in absolute intensity is measured. TreePol, on the other hand, is based on the interaction of a sample with unpolarized light and measures the polarization after interaction. After conversion as described in Section \ref{sec:CPCD}, both methods give the same quantitative results and show that in fresh leaves the fractional amount of differentially absorbed circularly polarized light is essentially the same as the amount of fractionally circular polarization by the chiral photosystems (i.e., in transmission). 

Figure \ref{fig:Spec} \textbf{A} \& \textbf{B} show that the in vivo circular polarization spectra for freshly plucked leaves are very similar to the typical spectra of isolated intact thylakoid membranes, similar to what was shown recently using CD \cite{Toth2016}. For freshly plucked leaves we observe a large positive circular dichroism band around 690 nm and a negative circular dichroism band around 660 nm. Both bands decrease in magnitude over time, but the positive band shows a faster decrease in magnitude for both the leaves stored in daylight \textbf{A} and in the dark \textbf{B} than the negative band; with a relatively large decrease after 7 days.
After 10 days of storage, a significant difference in the magnitude of the 690 nm band was observed between the spectra of the leaves stored in daylight and those stored in the dark ($49\pm  8.1\%$ vs. $69\pm 7.7 \%$ of the original magnitude). The difference in magnitude, albeit smaller, was also observed at 14 days ($15\pm  2.2\%$ vs. $26\pm 2.6 \%$ of the original magnitude) and 22 days ($9.6\pm  2.1\%$ vs. $18\pm 1.4 \%$ of the original magnitude). After 10 days, the negative band also showed a significant decrease in magnitude, but there was no significant difference between the storage conditions. The change in magnitude of the positive $V/I$ peak over time, averaged over 685-692 nm, is shown in Figure \ref{fig:Day} \textbf{A}. The sum of the absolute $V/I$ signal over 525-750 nm is shown in Figure \ref{fig:Day} \textbf{C}.

\subsection{Transmittance}
The transmittance spectra over time corresponding to the V/I measurements for the daylight and in the dark stored leaves are shown in figure \ref{fig:Spec} \textbf{C} and \textbf{D}. The transmittance decreases over time, mainly as a result of cell water loss. No significant differences in transmittance were observed between both storage conditions with the exception of the measurements at 7 days.

\subsection{Absorbance and fluorescence of chlorophyll extracts}
Figures \ref{fig:Spec} \textbf{E} and \textbf{F} show the absorbance spectra over time of the leaf chlorophyll extracts. Comparison of the extracts with the chlorophyll \textit{a} standard (data not shown) confirms that indeed chlorophyll \textit{a} is the major pigment extracted. Figures \ref{fig:Spec} \textbf{G} and \textbf{H} show the fluorescence spectra of the same extract, excited at 434 nm (Soret peak). Both fluorescence and absorbance measurements showed the same trend. In order to minimize the possible effects of solid constituents and other pigments, fluorescence data were used to calculate the chlorophyll \textit{a} concentrations, which are shown in Figure \ref{fig:Day} \textbf{B}. For both the daylight and the dark stored leaves the absorbance/fluorescence properties and hence the chlorophyll \textit{a} concentrations decrease over time. This decrease, however, is much more pronounced in the daylight stored leaves.

\begin{figure*}[]
  \centering
  \includegraphics[width=\linewidth]{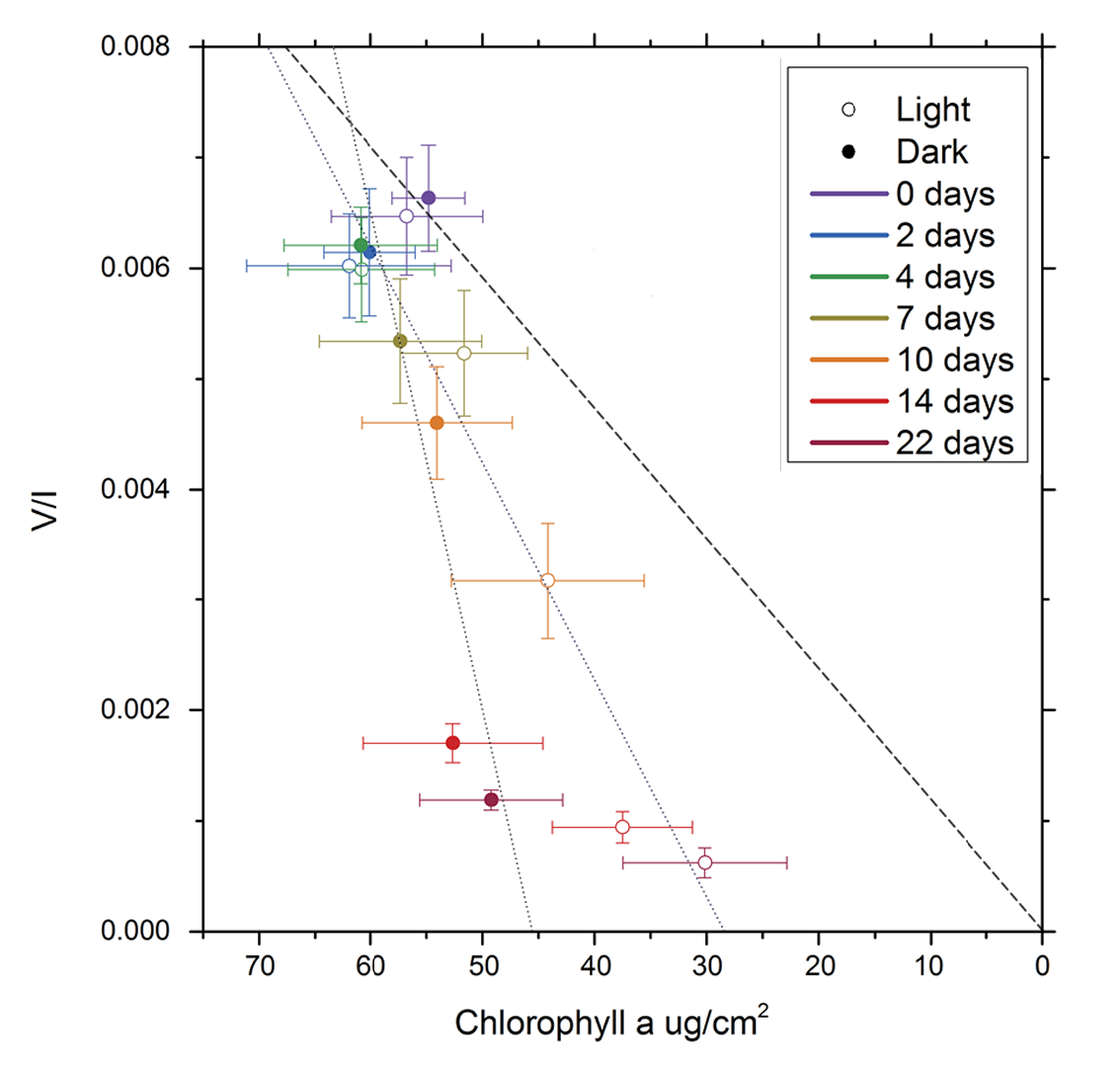}
  \caption{Average of the $V/I$ peak at 685-692 nm versus the Chlorophyll \textit{a} concentration over time for the daylight and dark stored leaves; the dashed line portrays a linear relationship. The dotted lines show the linear fit through the actual data for the daylight ($r^{2}=0.94$) and dark ($r^{2}=0.67$) stored leaves. The larger decrease in $V/I$ is primarily caused by the breakdown of the supramolecular structure; the chlorophyll \textit{a} levels show a much smaller decrease.  Vertical error bars denote the SE (n=6), horizontal error bars denote the SD (n=3).}
  \label{fig:ChlaVI}
\end{figure*}

\subsection{Chlorophyll \textit{a} concentrations vs V/I over time}
A plot of the changes in magnitude of the $V/I$ peak as a function of the chlorophyll concentrations is shown in Figure \ref{fig:ChlaVI}. Even though both $V/I$ and chlorophyll concentrations decrease over time, the decrease in $V/I$ is much larger and faster. This faster decrease becomes particularly apparent when one compares the data with the diagonal line, associated with  a hypothetical linear 1:1 relation between chlorophyll \textit{a} and $V/I$. The $V/I$ signal decreases faster in both daylight and dark stored leaves. For example, while the chlorophyll \textit{a} concentration per cm$^{2}$ of the 22 days old dark stored leaves is similar to that of fresh leaves, the signal of $V/I$ is less than 18\% of the original after 22 days.

\section{Discussion}

Detached leaves show a reduction in circular dichroism and polarization signal within days after storage. We have successfully demonstrated that circular polarization spectroscopy can be used in-vivo and post-mortem on leaves to measure the signal produced by the photosynthetic apparatus and we have shown that using circular polarization spectroscopy, healthy leaves can successfully be differentiated from senescing leaves. The similar results in both circular dichroism spectroscopy and circular polarization spectroscopy indicates isotropy in the fractional circular polarizing/absorbing component. As such there appear to be no systematic circular polarization differences between the adaxial and abaxial side of the leaf. 

\textit{H. helix}, is known for remaining green over prolonged periods of time under stress \cite{Warman1988, Horton1992}, which allowed us to carefully measure temporal changes. It is expected that the same phenomena can be observed for different species, but depending on the species, the effects are likely to occur within a shorter amount of time. 

Leaf senescence in general is a complex process with different underlying mechanisms. The effects of drought often includes the loss of chlorophylls, but this is not always the case \cite{Jaleel2009, Griffiths2014}. Here we observed that although the chlorophyll \textit{a} content decreases over time, the decrease in $V/I$ is much larger. While chlorophyll \textit{a} is at the basis of the $V/I$ signal, most of the signal results from the supramolecular chirality of the photosystems, which can be two orders of magnitude larger than the molecular signal \cite{Garab2009}. The absorbance maximum of the pigment coincides with the sign change in $V/I$, with a negative band at shorter wavelengths and a positive band at longer wavelengths. Both the negative and the positive band decreased in magnitude, and while the positive band decreased faster in magnitude over time and showed a significant difference between the storage conditions, these storage conditions did not lead to significant differences in the decrease of the negative band. 

While the chlorophyll \textit{a} concentration per leaf area decreases in the daylight stored leaves (but only slightly for the dark stored leaves), a much larger decrease is observed in the measured $V/I$. We hypothesize that the observed decrease is a direct result of the deterioration of the supramolecular chirality, and thus organization, which ultimately is the source of the observed signals. The dark stored leaves, with an end chlorophyll \textit{a} concentration similar to that of the fresh leaves, underline that the observed decrease cannot be explained by only the chlorophyll concentrations. As the supramolecular organization plays an important role in the regulation of photosynthetic activity, we propose the use of circular dichroism spectropolarimetry as a new tool for assessing vegetation stress. 

It has been shown that the two bands of the $V/I$ spectrum discussed above result from the superposition of two relatively independent signals \cite{Finzi1989}. The negative band is preferentially associated with the stacking of the thylakoid membranes, whereas the positive band is mainly associated with the lateral organization of the chiral macrodomains \cite{Cseh2000, Dobrikova2003, Jajoo2012, Garab1991}. It has furthermore been shown that external factors can influence the chiral order of these systems individually; whereas the membrane stacking is mainly influenced by the charge screening of divalent cations on the membrane interface, the lateral organization is predominantly influenced by the osmotic potential (i.e. it is absent in hypotonic solutions) \cite{Garab1991}. Indeed, leaf senescing is evidently coupled with a change in cell osmosis, but it is unclear how the drought-induced hypertonic environment will influence the lateral organization. Our results strongly suggest that the lateral organization decreases before unstacking. 

Figures \ref{fig:Spec} and \ref{fig:Day} show a significant difference in the decrease of the positive band between the leaves stored in the dark or in daylight (while no differences are observed in relative weight over time). While light-induced stress often leads to the loss of chlorophyll and the disruption of the photosystems, these effects only occur at very high light intensities well beyond the values used in our study \cite{Okada1992}. At low light levels senescing is often delayed compared to dark stored leaves \cite{Okada1992}. It should be noted, however, that the dark stored leaves were illuminated upon measurement and during weighing, which might be enough to delay senescing. It may well be that the combination of light and drought stress results in radicals influencing both the chlorophyll \textit{a} content and the supramolecular organization. 

Interestingly, it can be observed that the chlorophyll \textit{a} concentration per surface area does not decrease directly after detachment. In Figure \ref{fig:Spec} \textbf{E}, \textbf{F}, \textbf{G} and \textbf{H} it can be observed that both fluorescence and absorbance follow this trend. It is likely that these observations do not indicate an increase in chlorophyll \textit{a} per cell, but rather indicate cell contractions caused by the decrease in water content. These contractions might have influenced the observed $V/I$ spectra in the first few days after detachment, but this is less likely due to the normalization over $I$ (i.e., the measured intensity of $V$ is divided by the measured absolute intensity $I$).

Leaf transmittance in the near infrared does also decrease over time, but does not relate to the decrease of the $V/I$ bands. Furthermore, at the wavelengths around the $V/I$ positive maximum (685-692 nm) this effect is less pronounced and after 22 days the decrease was only significant for the daylight-stored leaves ($60 \pm 12\%$ of the original). The transmittance of the dark stored leaves around the $V/I$ maximum was never significantly different from the fresh leaves and was $100 \pm 21 \%$ of the original after 22 days. Leaf transmittance at longer wavelengths does decrease over time, but does not relate to the decrease of the $V/I$ bands. 
 
Large amounts of internal scattering could in principle lead to (de-)polarization effects, but with the large absorbance and low transmittance around the wavelengths of interest in our experiment we expect that this effect is minimal (multiple scattering events would be required to redirect the light towards the detector). Further support for this conclusion is provided by measurements on the commercial dichrograph, which showed a similar decrease in circular dichroism. In the case of the latter, the differential response to left- and right-handed circularly polarized light is measured. As such, multiple internal scattering events would be required before the light would react with the photosystems, which is highly implausible.

Finally, we want to emphasize the significance of circular polarization as both a valuable remotely applicable tool for vegetation monitoring on Earth and as remotely accessible means of detecting the presence of extraterrestrial life. We have successfully demonstrated that healthy leaves can be distinguished from unhealthy/dying leaves by the strong signal in the $V/I$ spectra. This signal, unique to vegetation, rapidly decreases over time if the chiral macrostructures are not actively maintained. 

Although this study was performed in transmittance mode, preliminary results show that the observations in reflectance mode give similar results. Future work will include measurements both in the laboratory as well as in the field.

\section{Acknowledgments}
Dr Wilfred R{\"o}ling, associate professor at the department of Molecular Cell Biology and initiator of this project, unexpectedly passed away on Friday the 25th of September 2015 at the age of 48 years. Our thoughts wander to the many moments of inspiration, joy and discovery we shared. Wilfred was passionate about science and sought to connect different fields together in order to approach it. We will continue to travel the avenues that he has outlined and we will miss him deeply. 

This work was supported by the Planetary and Exoplanetary Science Programme (PEPSci), grant 648.001.004, of the Netherlands Organisation for Scientific Research (NWO). 

We acknowledge Roberta Croce for the use of the CD spectrometer.
\section*{References}
\bibliography{mybibfile}

\end{document}